# Doping transition-metal atoms in graphene for atomic-scale tailoring of electronic, magnetic, and quantum topological properties

*Ondrej Dyck,[1][1] Lizhi Zhang,[2] Mina Yoon,[1] Jacob L. Swett,[3] Dale Hensley,[1] Cheng Zhang,[4] Philip D. Rack,[1,4] Jason D. Fowlkes,[1,4] Andrew R. Lupini,[1] Stephen Jesse[1]*

[1] *Center for Nanophase Materials Sciences, Oak Ridge National Laboratory, Oak Ridge, TN*

[2] *Department of Physics and Astronomy, University of Tennessee, Knoxville, TN*

[3] *University of Oxford, Department of Materials, Oxford OX1 3PH, UK*

[4] *Department of Materials Science and Engineering, University of Tennessee, Knoxville, TN*

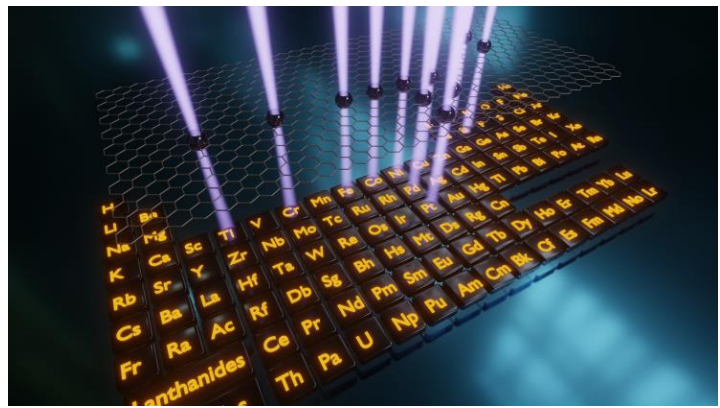

**Abstract**

Atomic-scale fabrication is an outstanding challenge and overarching goal for the nanoscience community. The practical implementation of moving and fixing atoms to a structure is non-trivial considering that one must spatially address the positioning of single atoms, provide a stabilizing scaffold to hold structures in place, and understand the details of their chemical bonding. Free-standing graphene offers a simplified platform for the development of atomic-scale fabrication and the focused electron beam in a scanning transmission electron microscope can be used to locally induce defects and sculpt the graphene. In this scenario, the graphene forms the stabilizing scaffold and the experimental question is whether a range of dopant atoms can be attached and incorporated into the lattice using a single technique and, from a theoretical perspective, we would like to know which dopants will create technologically interesting properties. Here, we demonstrate that the electron beam can be used to selectively and precisely insert a variety of transition metal atoms into graphene with highly localized control over the doping locations. We use first-principles density functional theory calculations with direct observation of the created structures to reveal the energetics of incorporating metal atoms into graphene and their magnetic, electronic, and quantum topological properties.

---

[1] Corresponding author. Tel: 336 262-7228. E-mail: dyckoe@ornl.gov (Ondrej Dyck)

**Introduction**

Graphene exhibits a number of unique properties of scientific interest for fundamental physics and applications. Not only does it exhibit high electron mobility,[1] surprising mechanical strength, and is optically transparent, but subtle details with regard to its electronic properties have made graphene a playground to explore fundamental physics.[2] Electron transport in graphene is governed by the relativistic Dirac equation, which leads to unusual electronic behavior such as the anomalous quantum hall effect and Klein tunneling.[3-7] These unique properties have prompted investigations to leverage graphene for quantum-based devices.[8-17] Proposed device designs require that the pristine graphene lattice be altered in highly specific ways to obtain the exact structure of interest; however, fabrication of such precision structures remains elusive. It has been proposed that the scanning transmission electron microscope (STEM) can be used as a fabrication platform by using the focused electron beam (e-beam) to tailor materials at the atomic scale[18] and a number of experimental examples suggest that such an idea is possible.[19] The e-beam can be used to introduce and evolve defects[20-25] as well as section fine structures such as graphene nanoribbons.[26-29] These capabilities demonstrate that fabrication workflows can be developed for the creation of graphene-based devices, which can be tailored to the single nanometer scale and possibly below. Many desired properties, however, are not intrinsic to carbon-only structures and necessitate the attachment or incorporation of foreign atoms. Significant efforts have been undertaken in theoretical publications predicting novel properties.[12, 30-33] Concurrently, experimental demonstrations have shown that graphene can host a variety of elements as dopants[34-42] or edge-passivating species[43-46] where doping of the graphene was performed *ex situ* using macro-scale techniques that did not permit spatial control. Small clusters of atoms have been shown to be separable from larger nanoparticles using the e-beam *in situ* and have been observed to move across various carbon-based substrates under the influence of the e-beam.[47] This suggests that separation and spatial control over the positioning of single atoms may be possible. Here, we experimentally demonstrate the locally controlled insertion of a wide variety of transition metal atoms into the graphene lattice using the e-beam as a manipulation tool[48, 49] and extend the theoretical treatment to include a larger portion of the periodic table. These nascent developments highlight a possible path toward atomic-scale patterning and when combined with recent progress in automated image analysis,[50-55] rapid accelerating voltage change capabilities,[56] and STEM-compatible *operando* device platforms,[57] can transform the STEM into a powerful atomic-scale fabrication instrument, as outlined by Kalinin et al.[18] First-principles density functional theory (DFT) calculations reveal metal-doped graphene with versatile magnetic, electronic, and non-trivial topological band structures, suggest them as potential building blocks for electronic and spintronic devices.

1. **Experimental Strategy**

Figure 1 shows a conceptual schematic illustrating the dopant insertion process. High-energy electrons from the e-beam are localized onto the graphene lattice where dopant insertions are desired. Incident electrons interacting with the atomic nuclei can impart a substantial energy transfer such that atoms can be ejected from the host (graphene) lattice leaving a point defect with dangling bonds. If foreign atoms are available and can diffuse across the graphene surface, they can bind to the defect site and passivate the dangling bonds. While ejecting carbon atoms from the graphene lattice is straight-forward using the e-beam, supplying foreign atoms for incorporation into the lattice is much more challenging. Three distinct strategies for inserting Si dopants, which are present on the surface of many graphene samples,[58] were previously experimentally demonstrated:[48, 49] (1) eject lattice atoms with a stationary e-beam as shown in Figure 1a-b followed by e-beam-induced sputtering of nearby source material on the graphene surface; (2) scan across a region of the sample that contains both clean graphene and the source material so that defects are created simultaneously with sputtering of the adatoms; (3) manually move the e-beam ("beam dragging") from the source material across the clean graphene to sputter atoms onto the graphene surface while creating defect locations for them to attach to.

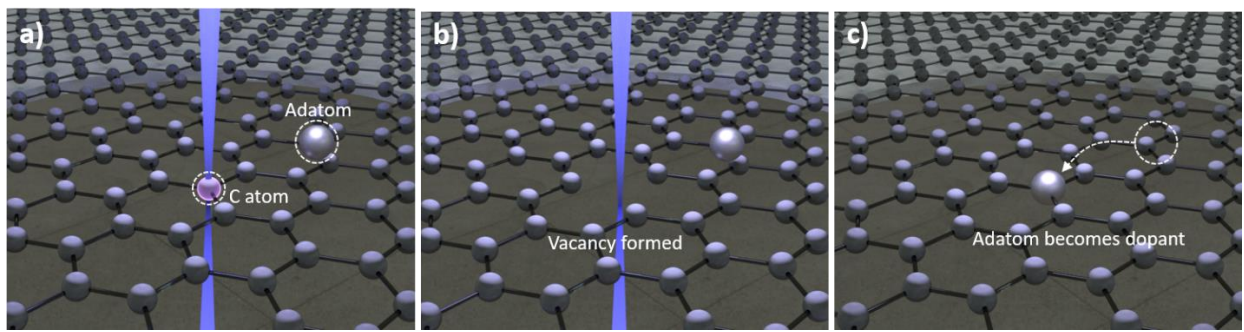

**Figure 1 Schematic of e-beam dopant insertion process.** a) The e-beam is positioned on a C atom in the graphene lattice. b) The e-beam delivers sufficient energy to knock out the C atom leaving a vacancy or other point defect. c) The e-beam is removed to prevent additional damage to the graphene lattice; adatoms on the graphene surface can migrate to the defect location at room temperature and bond to the lattice.

Strategies for dopant insertion were initially developed using Si as the dopant, primarily because Si is commonly found as a contaminant. Here, we intentionally introduce foreign atoms onto the graphene surfaces using *ex situ* e-beam evaporation and insert the foreign atoms into the graphene lattice *in situ* using the e-beam in the STEM. We note that the inherent Si contaminants are also present with the evaporated elements. Sample preparation details can be found in the Methods section.

## 2. Experimental Results

*2.1  Initial Demonstration*

As a first example to demonstrate several aspects of the doping process, a small amount of Ni was e-beam evaporated onto a graphene sample. Figure 2a and 2b show representative high angle annular dark field (HAADF) STEM images of the graphene plus evaporated Ni sample at low and high magnifications, respectively. Ni islands of ~3-5 nm diameter are observed on the graphene surface with many single Ni atoms dispersed between the Ni islands. The Ni islands were identified as metallic Ni using electron energy loss spectroscopy (EELS) as shown in Figure 2c, where the Ni $L_{23}$ characteristic edge is observed.

To demonstrate the dopant beam dragging insertion technique, we initially selected an area of pristine graphene near a Ni island, as shown in Figure 2d. In the present work, a beam energy of 100 keV is used, along with a low beam current (a few picoamps), which directly creates defects in the graphene, but at a slow rate. The stationary 100 kV e-beam was initially positioned on the Ni island, as indicated by the end of the red arrow. The e-beam was then directed or "dragged" from the Ni island across the pristine graphene, which serves to both sputter Ni atoms from the Ni island region and produce point defects in the adjacent graphene lattice creating attachment sites for the locally sputtered Ni atoms. Without the availability of attachment sites for the Ni adatoms, there is a low threshold for Ni mobility (0.23 eV).[59] Pixel dwell times on the order of 10 μs with a beam current of 20 pA produce on the order of $10^3$ electron impacts per pixel; resolving a single atom in a STEM image requires a residence time of > 100 μs or several tens of pixel acquisitions.

The 100 kV e-beam can transfer several eV directly to a Ni adatom and up to ~20 eV to the underlying C lattice; thus, capturing a STEM image of a Ni adatom is highly unlikely. Adatoms on the pristine area of the graphene will diffuse across the surface until they find other atoms at the edges to bond with, e.g., dangling bonds at defects. Brief interactions of diffusing adatoms with the e-beam at disparate pixel locations are indistinguishable from noise. Figure 2e shows the sample configuration after dragging the e-beam a few times from the Ni island onto the pristine graphene where five Ni atoms have extended from the edge of the Ni island (red dashed circle). A Fourier filtered image is shown inset with an atomic model overlay, which are used to capture and assess the positions of the dopants relative to the surrounding graphene lattice. The simulated overlay shows that the Ni atoms occupy four single graphene vacancies and one double vacancy. Figure 2f shows a higher magnification HAADF-STEM image of this same region after 180 s of scanning e-beam exposure; One of the Ni atoms stabilizes at the center of a restructured defect chain of 5-7 atom C rings, which is highlighted inset with an atomic structure overlay. The 5-atom rings are shown as yellow circles and the 7-atom rings are shown as blue circles. The Ni dopant occupies the intersection of two 7-atom rings.

Dopant insertion via simultaneous scanning of graphene and source material is illustrated in Figure 2g-i. In this example the e-beam was rastered across the Ni island and the adjacent pristine graphene. We observed a propensity of Ni atoms to attach to a close-by bilayer graphene step edge (left side of the images). As the same area continues to be scanned, more graphene defects are introduced and Ni atoms attach to the defects. The insets shown in Figure 2g and 2h are atomic model overlays of a 4-fold and 3-fold coordinated Ni dopant, respectively. As the number of defects in the graphene increases with additional scans, the defects in the graphene begin to coalesce to form larger holes, as shown in Figure 2i, and the Ni dopants begin to decorate the edges of the holes.

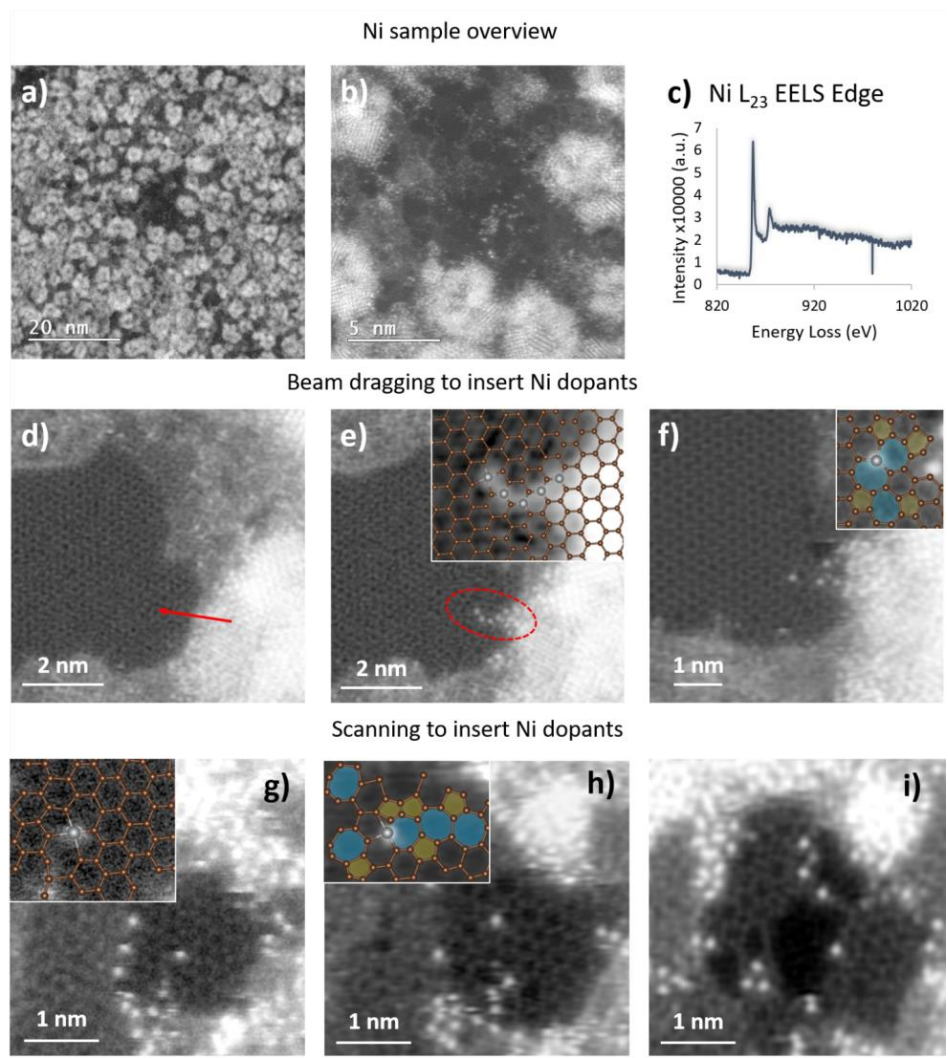

**Figure 2 Insertion of Ni atoms into the graphene lattice.** a-b) HAADF-STEM images show a sample overview at two different magnifications showing 3-5 nm Ni islands formed on the surface of graphene. Single Ni atoms are also observed between the Ni islands. c) Ni islands on the graphene surface as verified by Ni $L_{23}$ EELS core loss edge. d-e) Illustrate the beam dragging technique where the e-beam is positioned on the source material (tail of red arrow in d)) and dragged across the pristine graphene (head of red arrow in d)). This process sputters Ni source atoms while creating point defects in the graphene to which Ni atoms attach, e). Inset in e) shows Fourier filtered version of main HAADF-STEM image with atomic model overlay showing the positions of the Ni atoms. The Ni atom positions indicate occupation of single and di-vacancies. f) A higher resolution image of dopants after several minutes of e-beam exposure. An atomic model of the observed structure is overlaid. g)-i) An example of inserting Ni atoms by scanning the e-beam over a Ni island and pristine graphene. Initially the patch of graphene had no dopants; as defects are generated and Ni atoms are scattered onto graphene from the neighboring Ni island, Ni atoms attach to the defect sites and incorporate into the lattice. As more C atoms are knocked from lattice, holes begin to form and Ni atoms decorate the edges, i). Images e-f) and h-i) were filtered using principle component analysis in pycroscopy.[60, 61]

## 2.2   *Extending to Other Elements*

The above-described e-beam manipulation techniques have been extended to additional elements, demonstrating broader applications. Figure 3 and Figure 4 show examples of using these e-beam techniques to insert various elements into a graphene lattice; three example STEM images are shown with the specific element labeled on the left-hand side associated with each row of images. Associated EELS spectral fingerprints acquired on the larger metal nanoparticles/islands evaporated on the graphene surface are shown on the right. Red arrows in Figure 4 indicate the beam dragging locations and directions. Yellow markers distinguish the single metal atoms from residual Si dopants. Selected atomic models are shown inset as overlays on several of the STEM images. Image filtering was applied to selected images to improve the visibility of fine details, as described in the captions. Where filtering was deemed unnecessary for distinguishing such features the raw data is shown. For completeness we include results for inserting/moving Si atoms, which are present as a spurious contaminant on all our graphene samples. Figure 3 a-c show HAADF-STEM images of Si atoms inserted into the graphene lattice by scanning with a 100 kV e-beam. The Si EELS spectrum shown in 3 d was acquired at 60 kV on a single Si dopant; Si nanoparticles/islands were not present since Si was not evaporated on the surface as was done for other metal dopants but is a ubiquitous contaminant across the graphene. Figure 3 e-g show HAADF-STEM images of Ti atoms inserted into the graphene lattice by scanning a 100 kV e-beam. Si atoms were simultaneously inserted into the graphene lattice during the process and appear brighter than the C atoms but dimmer than the heavier metal atoms of interest in the HAADF-STEM images. Since both Si and the metal atoms are stabilized by defects in the graphene lattice, we frequently observed them bonded together or in close proximity to each other. The insets shown in Figure 3 e and 3 f include atomic model overlays of the observed structures. Close examination of the HAADF-STEM images reveals that the atoms shown in red in the models are slightly brighter than the C atoms and are likely O atoms, as O was also detected in the source material using EELS (not shown) and readily bonds to Ti. Figure 3 i-k show medium angle annular dark field (MAADF)-STEM images of Cr dopants inserted into the graphene lattice and Figure 3 m-s show examples of Fe dopants inserted into the lattice. Figure 3 q-s show the corresponding atomic model overlays with the five member rings (colored yellow) and the seven member rings (colored blue). For this example, the Fe was introduced onto the graphene surface through use of an iron cloride etchant used to remove the Cu foil that the graphene was grown on (as opposed to e-beam evaporation).

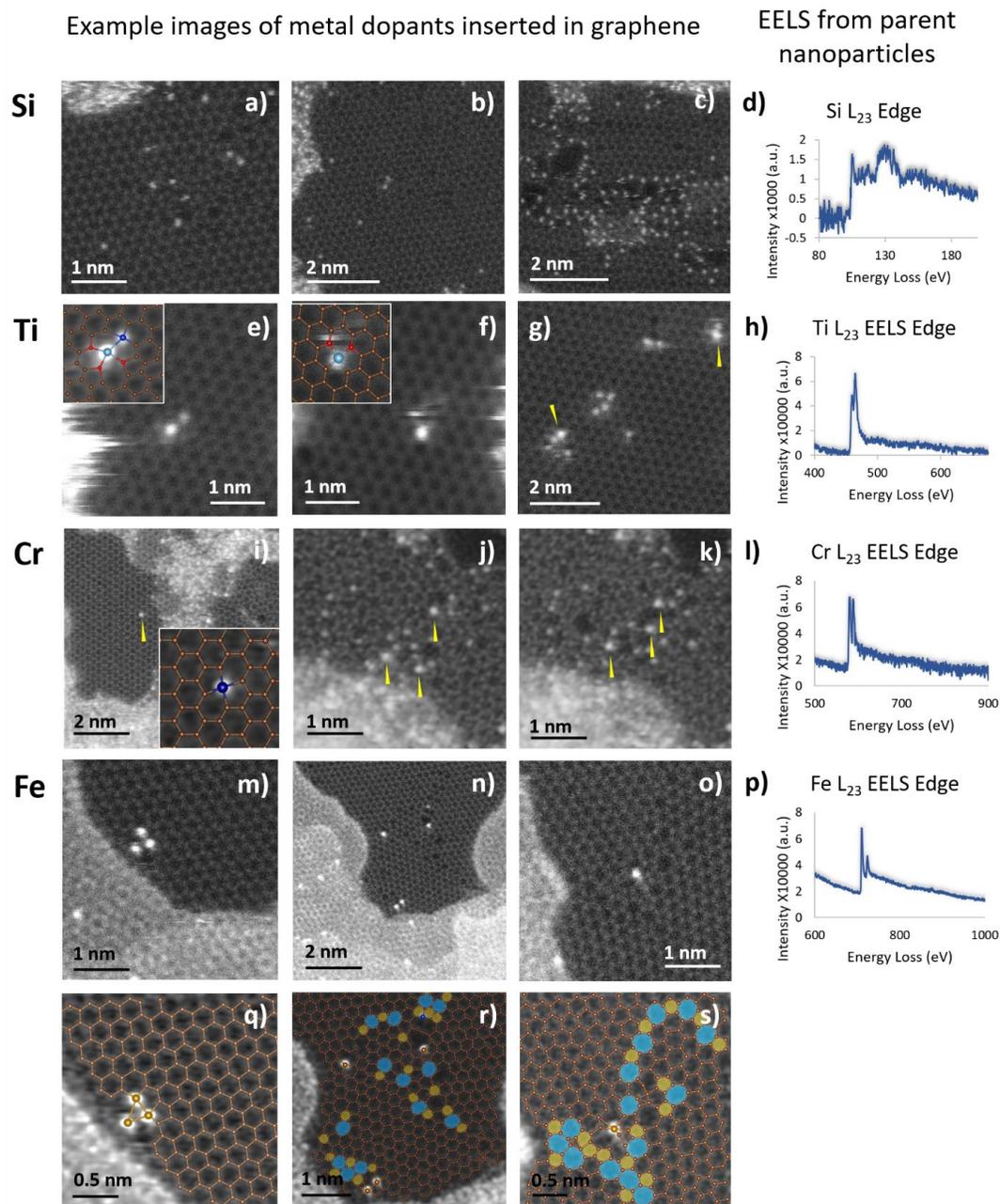

**Figure 3 Examples of metal atoms inserted into the graphene lattice.** The element inserted is labeled on the left-hand side of each row. Three HAADF-STEM images are shown for each element with an EELS spectrum acquired on the larger nanoparticles/islands that were deposited on the graphene surface. In each case, with the exception of the Fe sample, Si atoms were also present in the contamination and were simultaneously inserted during the process. Brighter atoms are metal atoms (highlighted by arrows). Insets show atomic models overlaid on the image with five-member rings (yellow) and seven-member rings (blue) highlighted. For Fe examples, (m)-(o), atomic models, (q)-(s), are shown in their own panels. Images (e)-(g) were acquired at 60 kV, the others were acquired at 100 kV. Images (a)-(c), (e)-(g), and (m)-(o) are unfiltered raw data, while the rest have been cleaned using image cleaning tools in pycroscopy.[60, 61]

Figure 4 shows examples for Co, Cu, Pd, Ag, and Pt. Figure 4 a-c show MAADF-STEM images of a beam-dragging sequence used to attach Co atoms to the graphene lattice at 100 kV accelerating voltage. Figure 4 a shows the initial configuration with the beam positioned at the origin of the red arrow and moved from the Co nanoparticle/island onto the graphene over several dragging iterations. The MAADF-STEM image in Figure 4 b was captured when the first atom attached to a defect. Finally, in Figure 4 c many Co dopants become attached to the graphene lattice as they migrate from the parent Co nanoparticle/island. Figure 4 e-g show MAADF-STEM images of Cu dopants inserted into the graphene lattice by simultaneously scanning over the deposited Cu parent nanoparticle/island and clean areas of the graphene. Figure 4i-k show a series of MAADF-STEM images of Pd dopants being inserted into the graphene lattice by beam-dragging. Figure 4 m-o show MAADF-STEM images of Ag dopants inserted into the graphene lattice at 100 kV. In Figure 4 n and 4 o, atomic model overlays are inset and show five member rings (yellow) and seven member rings (blue). Figure 4 q-s show HAADF-STEM images of Pt dopants inserted into the graphene at 100 kV. The Pt was deposited on the sample surface using e-beam-induced deposition from a metal-organic precursor gas source (as opposed to using e-beam evaporation for the other elements). The vertical line shown on the EELS spectrum in Figure 4 t indicates the edge onset to distinguish it from the bulk plasmon.

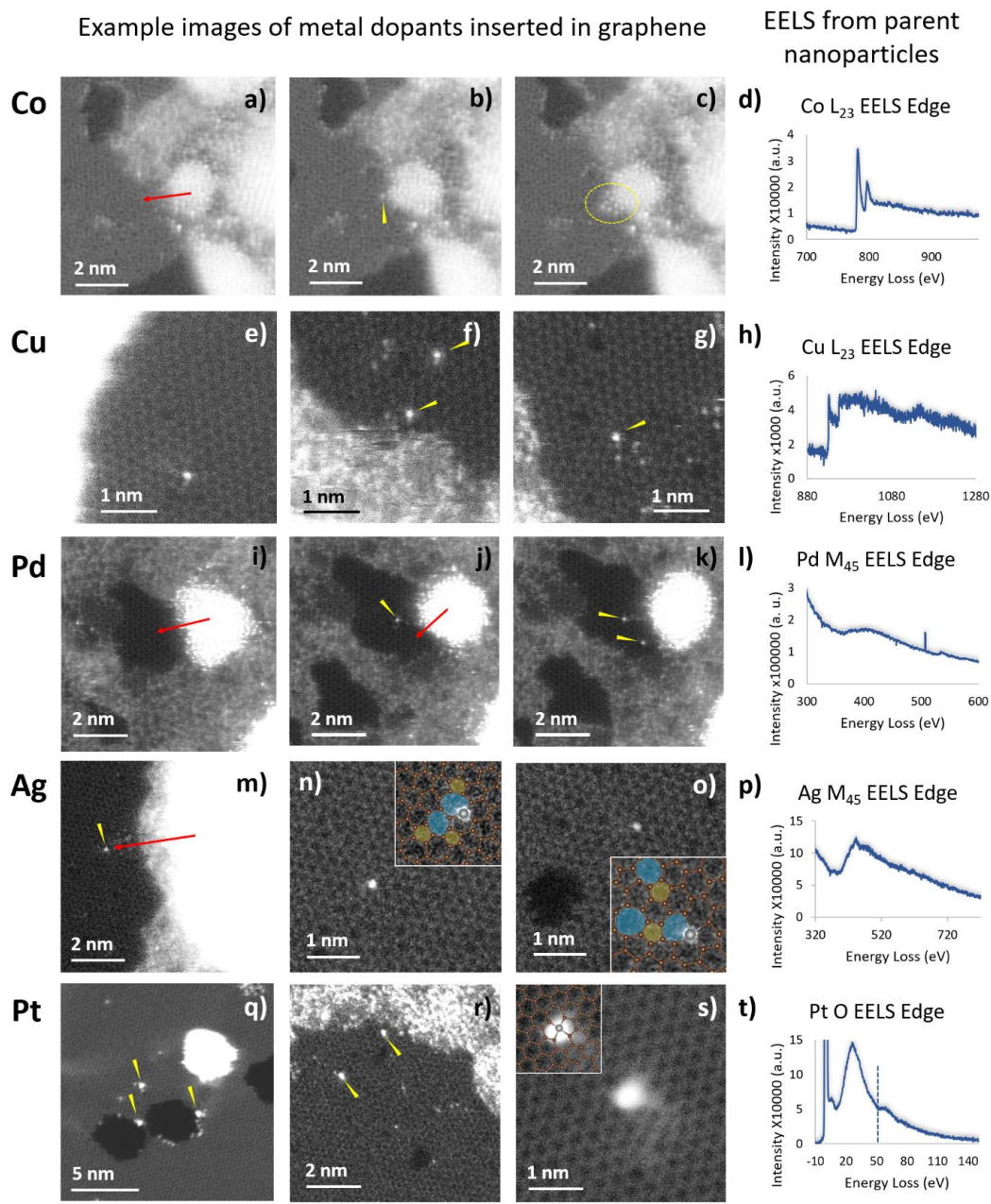

**Figure 4** . **Examples of metal atoms inserted into graphene.** The element inserted is labeled on the left-hand side of each row. Three STEM images and associated EELS spectrum acquired on the parent nanoparticle/island are shown. (a)-(c) shows a beam-dragging series for the insertion of Co. The initial sample configuration shown in a). The red arrow indicates where the e-beam was positioned on a Co nanoparticle/island and dragged across the graphene. b) The first Co atom attached to the lattice and c) after continuing to use the beam-dragging technique more Co atoms are attached to the lattice (yellow circle). For Pd, (i)-(k), depict a beam-dragging series where overlaid red arrows indicate the location where the beam was dragged. m) Insertion of Ag using beam dragging. The red arrow indicates the location of beam-dragging used to insert the Ag dopant. Images q) and s) were acquired at 60 kV and the rest were acquired at 100 kV. Images (e)-(g), (i)-(k), m), q), and s) are unfiltered raw data, while the rest have been cleaned using the image cleaning tools in pycroscopy.[60, 61]

## 2.3  *Experimental Discussion*

The examples shown in Figure 3 and 4 illustrate that the strategy for attaching dopant atoms to defect sites in graphene, employed previously for inserting residual Si contaminants,[48, 49] can be applied more universally. Here, we discuss the possibility that the dopant atoms could be merely surface adatoms. We cannot directly observe the z-height or spatially visualize bonding. This leaves the possibility of atoms simply adsorbed on the surface rather than covalently bound in substitutional (e.g. at a vacancy) or substitutional-like (e.g. at a divacancy) positions. Experimentally, we did not clearly resolve adatoms in sites such as bridging locations. This however does not mean that they do not occur, just that they were not clearly observable. Observation requires extended stability through time while simultaneously being irradiated by the focused 100 kV e-beam (to capture an image). Since adatom migration on graphene has a low activation energy it is highly likely that the beam will impart enough energy to an adatom (even with the beam tails) such that it becomes mobile and can diffuse to a new location. There is nothing preventing this from happening continuously so that adatoms are simply not imageable, wich appears to be the case. It is also possible that atoms become bound to reconstructed defects (e.g. 555-777 or similar structures[62]) with a bond energy between that of a substitutional-type defect and a surface adatom. We do not have clear images of these cases either, which suggests that the imparted beam energy is high enough to disrupt these structures as well. Thus, every image where the dopant and surrounding structure is clearly resolved appears consistent with a substitutional or substitutional-like structure.

Here, we show the manipulation and insertion of Ti, Cr, Fe, Co, Ni, Cu, Pd, Ag, and Pt into graphene using a scanned or dragged focused e-beam that relies on the introduction of source material onto the graphene surface to position the desired element in close proximity to the region of interest. From a fabrication perspective the coating of nanoparticles/islands on the graphene surface may pose a significant obstacle to ultimately obtain a pristinely structured device unless the nanoparticles/isalnds can be removed through a subsequent processing step such as rapid heating.[63, 64] As observed in Figure 2 a, deposition of materials directly onto the graphene surface results in a significant portion of the graphene being covered by the deposited species. Spatially selective deposition can be accomplished through lithography or by using an aperture or shadow mask to define deposition sites, which would enable multiple elements to be deposited on spatially distinct regions of the specimen while simultaneously masking off pristine regions. One could also envision using a micromanipulator to bring the source material in close proximity of a specimen *in situ*. Likewise, introduction of source atoms via thermal evaporation *in situ* or from a gas source may be another means of supplying atoms to the sample surface to attach to e-beam generated defects. *In situ* thermal evaporation, for example, would enable the source material to be physically separate from the device being fabricated and could allow for some degree of localization. Such a strategy would enable the combination of large area clean graphene with insertion of dopants at e-beam-generated defect sites. This capability, coupled with recent progress in *operando* STEM-compatible device platforms,[57] would enable high-precision devices to be fabricated and tested *in situ,* allowing for direct correlation between atomic structure and electronic properties.

The question remains, which atomic structures and species are technologically relevant, interesting, or useful. Performing an exhaustive experimental exploration of each possible dopant type and configuration is clearly prohibitive. An exhaustive theoretical approach is similarly intractable due to the extremely large number of possible configurations, especially when one begins to consider combinations of different elemental species. As a way forward, in the next section, we target the theoretical treatment of two of the simplest observed structures, a metal atom occupying a mono- and di-vacancy. This limits the simulated configurations to those that have been experimentally observed and enables a more rapid exploration to identify technologically relevant physical properties that may be amenable to fabrication using the e-beam techniques described here.

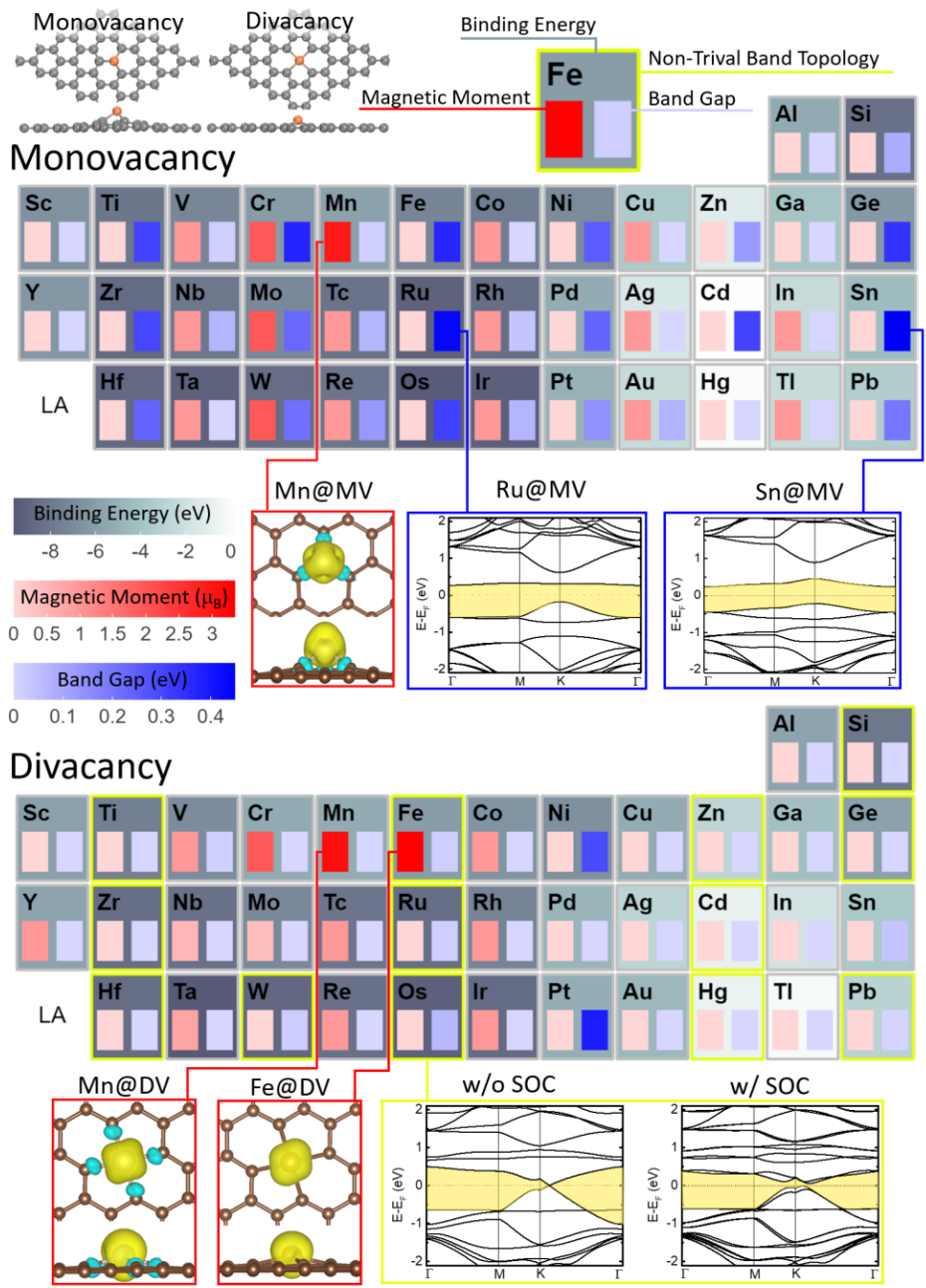

**Figure 5 First-principles modeling of metal-incorporated graphene.** The atomic configuration of metal-doped graphene with monovacancy (MV) and divacancy (DV) carbon atoms is show at the top. Metal binding energies, energy band gaps, and total magnetic moments are shown as colored swatches on the periodic table as indicated by the key at the top of the figure. Colorbars indicate the calculated values. Yellow outlines on the DV table indicate non-

trivial band topology. Spin density isosurfaces are shown for elements exhibiting the highest magnetic moments (Mn@MV, Mn@DV, and Fe@DV). Band structures are plotted for elements exhibiting the largest band gaps (Ru@MV and Sn@MV). An example of the non-trivial band topology is shown for Os@DV where spin-orbit coupling (SOC) results in an emergent energy gap at the Dirac point indicating it is a topological insulator.

## 3. Theoretical Results

Having experimentally demonstrated the insertion of a wide range of metal atoms into a graphene lattice using an e-beam, we turn to first-principles density functional theory (DFT) calculations to reveal the metal-binding energies, electronic, and magnetic properties of dopants in graphene monovacancies (MV) and divacancies (DV) for a large portion of the periodic table (see Methods section for details). We model the prinstine graphene using 5x5 supercells containing 50 carbon atoms and metal-doped graphene with two types of doping configurations, as shown at the top of Figure 5. Metal atoms are incorporated into either a MV site or DV site, resulting in ~2% doping concentrations. We chose all the transition metals of periods 4-6 in the periodic table as well as a few post-transition metals and other metals. The complete list of metals is shown in Figure 5. The sizes of the metal atoms considered are much larger than that of a carbon atom; thus a metal-dopant attached to the MV site protrudes from the basal plane and forms $sp^3$ bonds with carbon atoms, while metals at the DV site with a slighly larger hole stay closer to the basal plane with only a slight protrusion, as shown at the top of Figure 5. The competition between the local strain relaxation and the formation of strong $sp^3$ bonds dictates the stability of the metal atoms at the divacancy and monovacancy sites. We confirm that incorporating metals into the defect sites is always an exothermic process. Figure 5 compares the metal binding energies ($E_b$) of metal-incorporated graphene with $E_b$ defined as $E_{tot} - E_M - E_{def-grap}$, where $E_{tot}$, $E_M$, and $E_{def-grap}$ are the total energies of metal-doped graphene, a metal atom (M), and defective graphene with one or two carbon vacancies, respectively. Metal binding is favorable for all the considered cases with respect to the system of graphene with defects and metals as a gas phase; the metal binding can be as high as ~9 eV per atom as indicated on the gray binding energy colorbar. The binding energies of transition-metal atoms oscillate with their *d*-orbital occupancy and have the weakest bindings for Zn, Cd, and Hg. For example, for period 4 transition metals, 3*d* orbitals start being occupied from Sc and gradually increase with the atomic number, achieving full occupancy with Zn, which also has a fully occupied 4*s* orbital. The energy gain to include Sc at the vacancy sites is ~6 eV, that increases further with Ti and slightly decreases as the occupancy increases to Mn, which has majority spin fully occupying the 3*d* orbitals and fully occupied 4*s* orbitals. The binding energy becomes stronger as the minority spin components begin occupying the 3*d* orbitals up to Ni, which has half of the minority spin occupying the 3*d* orbitals. The binding energy then substantially weakens for Cu and exhibits a local minimum at Zn with fully occupied 3*d* and 4*s* orbitals. This periodically oscillating binding energy pattern repeats for the other periods of transition metals with slightly stronger binding energies at the beginning of the oscillation and slightly weaker binding energies at the end. For example, moving down the periodic table from Zn to Hg results in a weaker binding energy but moving down the periodic table from Co to Ir results in a stronger binding energy. Figure 5 also presents the total magnetic moments of the metal-incorporated graphene at both the MV and DV sites, shown by the red swatches. The highest magnetic moment of ~3$\mu_B$ can be achieved with Mn at either a MV or DV site and Fe at a DV site, consistent with that reported in the literature.[30] Spin-density isosurfaces for these cases are plotted in Figure 5 with an isosurface value of 0.00015 e/Å$^3$.

Graphene electronic band structures can also be tuned by doping with metals; energy band gaps can be opened to ~0.4eV, shown by the blue swatches in Figure 5. Band diagrams are plotted for the two largest band gaps, Ru and Sn at a MV site. In addition, we identified that some dopants form Dirac bands with energy gaps at the Dirac points that open upon the application of spin-orbit coupling, which is an indication of non-trivial band topology. These dopants are Ti, Fe, Zn, Zr, Ru, Cd, Hf, W, Os, Hg, Si, Ge, and Pb that occupy the DV site, highlighted by yellow outlines in Figure 5. As an example, the band structure for Os is plotted with and without spin-orbit coupling showing the Dirac point (between K and Γ) at the Fermi level. Upon spin-orbit coupling (SOC), an energy gap opens by ~65.5 meV at the Dirac point, indicating a non-trivial band topology and proving a topological insulator. Tabulated values for binding energy, magnetic moment, and band gap are given in the supplementary information. Note that the discussed properties are based on our specific model system — a homogeneously metal-doped graphene (M@DV or M@MV) with ~2% doping and these properties may change significantly with dopant density and the symmetry of the supercell. More detailed theoretical exploration of these parameters will be addressed elsewhere. In summary, our theory predicts versatile electronic, magnetic, and quantum topological properties of metal-doped graphene, suggesting they may form interesting building blocks for topological electronic and spintronic applications.

## 4. Conclusion

We present a series of atomic manipulation experiments demonstrating the localized insertion of a wide variety of transition metal atoms in graphene, namely Ti, Cr, Fe, Co, Ni, Cu, Pd, Ag, and Pt. We comment on how these and related strategies can be leveraged for top-down fabrication of atomic and nanometer-scale devices to harness the unique properties that emerge at these length scales. Finally, we expand the knowledge base regarding the effect of doping graphene with all types of transition metals using first principles calculations to guide future attempts at tailoring the properties of graphene.

## 5. Methods

## 5.1 Sample Preparation

We began by transferring chemical vapor deposition (CVD)-grown graphene to TEM grids using a wet transfer method and baking in an Ar-$O_2$ (90%-10%) environment at 500 °C for 1.5 h to clean the graphene.[64] Samples were then transferred to an e-beam evaporator (Thermionics VE-240 e-beam evaporator operated at a pressure <5 x $10^{-6}$ Torr) where 1-10 Å of material from various elements were evaporated onto the graphene surface. A quartz crystal microbalance was used to estimate the nominal thickness of the deposited layer and a mechanical shutter was used to start and stop the deposition. Samples were then baked in vacuum at 160 °C for 8-10 hours before being imaged in a Nion UltraSTEM 200. The nominal beam current was 20 pA and convergence angle of 30 mrad.

## 5.2 DFT Details

All the calculations are based on first-principles DFT using the Vienna ab-initio simulation package (VASP)[65] with the projector augmented wave method and a generalized gradient approximation (GGA) in the form of Perdew–Burke–Ernzerhof is adopted for the exchange-correlation functional.[66] The energy cutoff of the plane-wave basis sets is 400 eV and a $5\times5\times1$ Monkhorst-Park k-point mesh is used for the $5\times5\times1$ supercell calculation. Supercell structures include a 10 Å vacuum layer and all structures are fully relaxed until the residual forces on each atom are less than 0.01 eV/Å.

**Acknowledgement**

This material is based upon work supported by the U.S. Department of Energy, Office of Science, Basic Energy Sciences, Materials Sciences and Engineering Division (O.D., M.Y., A.R.L., S.J.) and Oak Ridge National Laboratory's Center for Nanophase Materials Sciences (CNMS), a U.S.

Department of Energy, Office of Science User Facility (D.H., L.Z.). P.D.R. and J.D.F. acknowledge support for the e-beam-induced deposition provided by the Nanofabrication Research Laboratory at CNMS. CZ acknowledges support from the US Department of Energy (DOE) under Grant No. DOE DE-SC0002136.

# References

1. Novoselov, K. S.; Geim, A. K.; Morozov, S. V.; Jiang, D.; Zhang, Y.; Dubonos, S. V.; Grigorieva, I. V.; Firsov, A. A., Electric Field Effect in Atomically Thin Carbon Films. *Science* **2004,** *306* (5696), 666-669.
2. Katsnelson, M. I.; Novoselov, K. S., Graphene: New bridge between condensed matter physics and quantum electrodynamics. *Solid State Commun.* **2007,** *143* (1), 3-13.
3. Bolotin, K. I.; Ghahari, F.; Shulman, M. D.; Stormer, H. L.; Kim, P., Observation of the fractional quantum Hall effect in graphene. *Nature* **2009,** *462* (7270), 196-199.
4. Novoselov, K. S.; Geim, A. K.; Morozov, S. V.; Jiang, D.; Katsnelson, M. I.; Grigorieva, I. V.; Dubonos, S. V.; Firsov, A. A., Two-dimensional gas of massless Dirac fermions in graphene. *Nature* **2005,** *438*, 197.
5. Beenakker, C. W. J., Colloquium: Andreev reflection and Klein tunneling in graphene. *Rev. Mod. Phys.* **2008,** *80* (4), 1337-1354.
6. Molitor, F.; Güttinger, J.; Stampfer, C.; Dröscher, S.; Jacobsen, A.; Ihn, T.; Ensslin, K., Electronic properties of graphene nanostructures. *J. Phys.: Condens. Matter* **2011,** *23* (24), 243201.
7. Castro Neto, A. H.; Guinea, F.; Peres, N. M. R.; Novoselov, K. S.; Geim, A. K., The electronic properties of graphene. *Rev. Mod. Phys.* **2009,** *81* (1), 109-162.
8. Patrik, R.; Björn, T., Quantum dots and spin qubits in graphene. *Nanotechnology* **2010,** *21* (30), 302001.
9. Trauzettel, B.; Bulaev, D. V.; Loss, D.; Burkard, G., Spin qubits in graphene quantum dots. *Nat. Phys.* **2007,** *3*, 192.
10. Han, W.; Kawakami, R. K.; Gmitra, M.; Fabian, J., Graphene spintronics. *Nat. Nanotechnol.* **2014,** *9*, 794.
11. Pesin, D.; MacDonald, A. H., Spintronics and pseudospintronics in graphene and topological insulators. *Nat. Mater.* **2012,** *11*, 409.
12. Da, H.; Feng, Y. P.; Liang, G., Transition-Metal-Atom-Embedded Graphane and Its Spintronic Device Applications. *J. Phys. Chem. C* **2011,** *115* (46), 22701-22706.
13. Wang, Z. F.; Li, Q.; Shi, Q. W.; Wang, X.; Hou, J. G.; Zheng, H.; Chen, J., Ballistic rectification in a Z-shaped graphene nanoribbon junction. *Appl. Phys. Lett.* **2008,** *92* (13), 133119.
14. Wang, Z. F.; Shi, Q. W.; Li, Q.; Wang, X.; Hou, J. G.; Zheng, H.; Yao, Y.; Chen, J., Z-shaped graphene nanoribbon quantum dot device. *Appl. Phys. Lett.* **2007,** *91* (5), 053109.
15. Saffarzadeh, A.; Farghadan, R., A spin-filter device based on armchair graphene nanoribbons. *Appl. Phys. Lett.* **2011,** *98* (2), 023106.
16. Zeng, M. G.; Shen, L.; Cai, Y. Q.; Sha, Z. D.; Feng, Y. P., Perfect spin-filter and spin-valve in carbon atomic chains. *Appl. Phys. Lett.* **2010,** *96* (4), 042104.
17. Fal'ko, V., Quantum information on chicken wire. *Nat. Phys.* **2007,** *3* (3), 151-152.

18. Kalinin, S. V.; Borisevich, A.; Jesse, S., Fire up the atom forge. *Nature* **2016,** *539* (7630), 485-487.
19. Dyck, O.; Ziatdinov, M.; Lingerfelt, D. B.; Unocic, R. R.; Hudak, B. M.; Lupini, A. R.; Jesse, S.; Kalinin, S. V., Atom-by-atom fabrication with electron beams. *Nat. Rev. Mater.* **2019**.
20. Kotakoski, J.; Mangler, C.; Meyer, J. C., Imaging atomic-level random walk of a point defect in graphene. *Nat. Commun.* **2014,** *5*, 3991.
21. Kotakoski, J.; Meyer, J. C.; Kurasch, S.; Santos-Cottin, D.; Kaiser, U.; Krasheninnikov, A. V., Stone-Wales-type transformations in carbon nanostructures driven by electron irradiation. *Phys. Rev. B* **2011,** *83* (24), 245420.
22. Robertson, A. W.; He, K.; Kirkland, A. I.; Warner, J. H., Inflating Graphene with Atomic Scale Blisters. *Nano Lett.* **2014,** *14* (2), 908-914.
23. Robertson, A. W.; Lee, G.-D.; He, K.; Yoon, E.; Kirkland, A. I.; Warner, J. H., The Role of the Bridging Atom in Stabilizing Odd Numbered Graphene Vacancies. *Nano Lett.* **2014,** *14* (7), 3972-3980.
24. Robertson, A. W.; Lee, G.-D.; He, K.; Yoon, E.; Kirkland, A. I.; Warner, J. H., Stability and Dynamics of the Tetravacancy in Graphene. *Nano Lett.* **2014,** *14* (3), 1634-1642.
25. Dyck, O.; Yoon, M.; Zhang, L.; Lupini, A. R.; Swett, J. L.; Jesse, S., Doping of Cr in Graphene Using Electron Beam Manipulation for Functional Defect Engineering. *ACS Applied Nano Materials* **2020**.
26. Rodríguez-Manzo, J. A.; Qi, Z. J.; Crook, A.; Ahn, J.-H.; Johnson, A. T. C.; Drndić, M., In Situ Transmission Electron Microscopy Modulation of Transport in Graphene Nanoribbons. *ACS Nano* **2016,** *10* (4), 4004-4010.
27. Qi, Z. J.; Rodríguez-Manzo, J. A.; Hong, S. J.; Park, Y. W.; Stach, E. A.; Drndić, M.; Johnson, A. T. C., *Direct electron beam patterning of sub-5nm monolayer graphene interconnects*. SPIE: 2013; Vol. 8680.
28. Fischbein, M. D.; Drndić, M., Electron beam nanosculpting of suspended graphene sheets. *Appl. Phys. Lett.* **2008,** *93* (11), 113107.
29. Qi, Z. J.; Rodríguez-Manzo, J. A.; Botello-Méndez, A. R.; Hong, S. J.; Stach, E. A.; Park, Y. W.; Charlier, J.-C.; Drndić, M.; Johnson, A. T. C., Correlating Atomic Structure and Transport in Suspended Graphene Nanoribbons. *Nano Lett.* **2014,** *14* (8), 4238-4244.
30. Krasheninnikov, A. V.; Lehtinen, P. O.; Foster, A. S.; Pyykkö, P.; Nieminen, R. M., Embedding Transition-Metal Atoms in Graphene: Structure, Bonding, and Magnetism. *Phys. Rev. Lett.* **2009,** *102* (12), 126807.
31. Rizzo, D. J.; Veber, G.; Cao, T.; Bronner, C.; Chen, T.; Zhao, F.; Rodriguez, H.; Louie, S. G.; Crommie, M. F.; Fischer, F. R., Topological band engineering of graphene nanoribbons. *Nature* **2018,** *560* (7717), 204-208.
32. Thakur, J.; Saini, H. S.; Singh, M.; Reshak, A. H.; Kashyap, M. K., Quest for magnetism in graphene via Cr- and Mo-doping: A DFT approach. *Physica E: Low-dimensional Systems and Nanostructures* **2016,** *78*, 35-40.
33. Gorjizadeh, N.; Farajian, A. A.; Esfarjani, K.; Kawazoe, Y., Spin and band-gap engineering in doped graphene nanoribbons. *Phys. Rev. B* **2008,** *78* (15), 155427.
34. Lin, Y.-C.; Teng, P.-Y.; Chiu, P.-W.; Suenaga, K., Exploring the Single Atom Spin State by Electron Spectroscopy. *Phys. Rev. Lett.* **2015,** *115* (20), 206803.
35. Tripathi, M.; Markevich, A.; Böttger, R.; Facsko, S.; Besley, E.; Kotakoski, J.; Susi, T., Implanting Germanium into Graphene. *ACS Nano* **2018,** *12* (5), 4641-4647.
36. Warner, J. H.; Lin, Y.-C.; He, K.; Koshino, M.; Suenaga, K., Stability and Spectroscopy of Single Nitrogen Dopants in Graphene at Elevated Temperatures. *ACS Nano* **2014,** *8* (11), 11806-11815.
37. Su, C.; Tripathi, M.; Yan, Q.-B.; Wang, Z.; Zhang, Z.; Hofer, C.; Wang, H.; Basile, L.; Su, G.; Dong, M.; Meyer, J. C.; Kotakoski, J.; Kong, J.; Idrobo, J.-C.; Susi, T.; Li, J., Engineering single-atom dynamics with electron irradiation. *Science Advances* **2019,** *5* (5), eaav2252.

38. Robertson, A. W.; Montanari, B.; He, K.; Kim, J.; Allen, C. S.; Wu, Y. A.; Olivier, J.; Neethling, J.; Harrison, N.; Kirkland, A. I.; Warner, J. H., Dynamics of Single Fe Atoms in Graphene Vacancies. *Nano Lett.* **2013,** *13* (4), 1468-1475.
39. He, Z.; He, K.; Robertson, A. W.; Kirkland, A. I.; Kim, D.; Ihm, J.; Yoon, E.; Lee, G.-D.; Warner, J. H., Atomic Structure and Dynamics of Metal Dopant Pairs in Graphene. *Nano Lett.* **2014,** *14* (7), 3766-3772.
40. Kepaptsoglou, D.; Hardcastle, T. P.; Seabourne, C. R.; Bangert, U.; Zan, R.; Amani, J. A.; Hofsäss, H.; Nicholls, R. J.; Brydson, R. M. D.; Scott, A. J.; Ramasse, Q. M., Electronic Structure Modification of Ion Implanted Graphene: The Spectroscopic Signatures of p- and n-Type Doping. *ACS Nano* **2015,** *9* (11), 11398-11407.
41. Zan, R.; Bangert, U.; Ramasse, Q.; Novoselov, K. S., Metal−Graphene Interaction Studied via Atomic Resolution Scanning Transmission Electron Microscopy. *Nano Lett.* **2011,** *11* (3), 1087-1092.
42. Ramasse, Q. M.; Seabourne, C. R.; Kepaptsoglou, D.-M.; Zan, R.; Bangert, U.; Scott, A. J., Probing the Bonding and Electronic Structure of Single Atom Dopants in Graphene with Electron Energy Loss Spectroscopy. *Nano Lett.* **2013,** *13* (10), 4989-4995.
43. Ta, H. Q.; Zhao, L.; Yin, W.; Pohl, D.; Rellinghaus, B.; Gemming, T.; Trzebicka, B.; Palisaitis, J.; Jing, G.; Persson, P. O. Å.; Liu, Z.; Bachmatiuk, A.; Rümmeli, M. H., Single Cr atom catalytic growth of graphene. *Nano Res.* **2017**.
44. Zhao, J.; Deng, Q.; Avdoshenko, S. M.; Fu, L.; Eckert, J.; Rümmeli, M. H., Direct in situ observations of single Fe atom catalytic processes and anomalous diffusion at graphene edges. *Proc. Natl. Acad. Sci. U. S. A.* **2014,** *111* (44), 15641-15646.
45. Wang, W. L.; Santos, E. J. G.; Jiang, B.; Cubuk, E. D.; Ophus, C.; Centeno, A.; Pesquera, A.; Zurutuza, A.; Ciston, J.; Westervelt, R.; Kaxiras, E., Direct Observation of a Long-Lived Single-Atom Catalyst Chiseling Atomic Structures in Graphene. *Nano Lett.* **2014,** *14* (2), 450-455.
46. Ramasse, Q. M.; Zan, R.; Bangert, U.; Boukhvalov, D. W.; Son, Y.-W.; Novoselov, K. S., Direct Experimental Evidence of Metal-Mediated Etching of Suspended Graphene. *ACS Nano* **2012,** *6* (5), 4063-4071.
47. Cretu, O.; Rodríguez-Manzo, J. A.; Demortière, A.; Banhart, F., Electron beam-induced formation and displacement of metal clusters on graphene, carbon nanotubes and amorphous carbon. *Carbon* **2012,** *50* (1), 259-264.
48. Dyck, O.; Kim, S.; Jimenez-Izal, E.; Alexandrova, A. N.; Kalinin, S. V.; Jesse, S., Building Structures Atom by Atom via Electron Beam Manipulation. *Small* **2018,** *14* (38), 1801771.
49. Dyck, O.; Kim, S.; Kalinin, S. V.; Jesse, S., Placing single atoms in graphene with a scanning transmission electron microscope. *Appl. Phys. Lett.* **2017,** *111* (11), 113104.
50. Maksov, A.; Dyck, O.; Wang, K.; Xiao, K.; Geohegan, D. B.; Sumpter, B. G.; Vasudevan, R. K.; Jesse, S.; Kalinin, S. V.; Ziatdinov, M., Deep learning analysis of defect and phase evolution during electron beam-induced transformations in WS2. *npj Comput. Mater.* **2019,** *5* (1), 12.
51. Ziatdinov, M.; Dyck, O.; Li, X.; Sumpter, B. G.; Jesse, S.; Vasudevan, R. K.; Kalinin, S. V., Building and exploring libraries of atomic defects in graphene: Scanning transmission electron and scanning tunneling microscopy study. *Science Advances* **2019,** *5* (9), 9.
52. Ziatdinov, M.; Dyck, O.; Jesse, S.; Kalinin, S. V., Atomic Mechanisms for the Si Atom Dynamics in Graphene: Chemical Transformations at the Edge and in the Bulk. *Adv. Funct. Mater.* **2019,** *29* (52), 1904480.
53. Vasudevan, R. K.; Laanait, N.; Ferragut, E. M.; Wang, K.; Geohegan, D. B.; Xiao, K.; Ziatdinov, M.; Jesse, S.; Dyck, O.; Kalinin, S. V., Mapping mesoscopic phase evolution during E-beam induced transformations via deep learning of atomically resolved images. *npj Comput. Mater.* **2018,** *4* (1), 30.
54. Ziatdinov, M.; Dyck, O.; Maksov, A.; Li, X.; Sang, X.; Xiao, K.; Unocic, R. R.; Vasudevan, R.; Jesse, S.; Kalinin, S. V., Deep Learning of Atomically Resolved Scanning Transmission Electron Microscopy

Images: Chemical Identification and Tracking Local Transformations. *ACS Nano* **2017,** *11* (12), 12742-12752.
55. Ziatdinov, M.; Maksov, A.; Kalinin, S. V., Learning surface molecular structures via machine vision. *npj Comput. Mater.* **2017,** *3* (1), 31.
56. Dyck, O.; Jesse, S.; Delby, N.; Kalinin, S. V.; Lupini, A. R., Variable voltage electron microscopy: Toward atom-by-atom fabrication in 2D materials. *Ultramicroscopy* **2020,** *211*, 112949.
57. Swett, J. L.; Kravchenko, I. I.; Dyck, O. E.; Jesse, S.; Mol, J. A., A Versatile Common Platform for Quantum Transport Measurements in Fluidic, Cryogenic, and In Situ Electron Microscopy Environments. *Microsc. Microanal.* **2019,** *25* (S2), 972-973.
58. Lisi, N.; Dikonimos, T.; Buonocore, F.; Pittori, M.; Mazzaro, R.; Rizzoli, R.; Marras, S.; Capasso, A., Contamination-free graphene by chemical vapor deposition in quartz furnaces. *Sci. Rep.* **2017,** *7* (1), 9927.
59. Manadé, M.; Viñes, F.; Illas, F., Transition metal adatoms on graphene: A systematic density functional study. *Carbon* **2015,** *95*, 525-534.
60. Somnath, S.; Smith, C. R.; Laanit, N.; Jesse, S. *Pycroscopy*, 0.59.7; Oak Ridge National Laboratory, 2016.
61. Somnath, S.; Smith, C. R.; Laanait, N.; Vasudevan, R. K.; Ievlev, A.; Belianinov, A.; Lupini, A. R.; Shankar, M.; Kalinin, S. V.; Jesse, S., USID and Pycroscopy--Open frameworks for storing and analyzing spectroscopic and imaging data. *arXiv preprint arXiv:1903.09515* **2019**.
62. Cretu, O.; Krasheninnikov, A. V.; Rodríguez-Manzo, J. A.; Sun, L.; Nieminen, R. M.; Banhart, F., Migration and Localization of Metal Atoms on Strained Graphene. *Phys. Rev. Lett.* **2010,** *105* (19), 196102.
63. Tripathi, M.; Mittelberger, A.; Mustonen, K.; Mangler, C.; Kotakoski, J.; Meyer, J. C.; Susi, T., Cleaning graphene: Comparing heat treatments in air and in vacuum. *physica status solidi (RRL) – Rapid Research Letters* **2017,** *11* (8), 1700124-n/a.
64. Dyck, O.; Kim, S.; Kalinin, S. V.; Jesse, S., Mitigating e-beam-induced hydrocarbon deposition on graphene for atomic-scale scanning transmission electron microscopy studies. *Journal of Vacuum Science & Technology, B: Nanotechnology & Microelectronics: Materials, Processing, Measurement, & Phenomena* **2017,** *36* (1), 011801.
65. Kresse, G.; Furthmüller, J., Efficient iterative schemes for ab initio total-energy calculations using a plane-wave basis set. *Phys. Rev. B* **1996,** *54* (16), 11169-11186.
66. Perdew, J. P.; Burke, K.; Ernzerhof, M., Generalized Gradient Approximation Made Simple. *Phys. Rev. Lett.* **1996,** *77* (18), 3865-3868.